\documentclass[proceedings]{stacs}
\stacsheading{2009}{231--242}{Freiburg}
\firstpageno{231}

\usepackage{times}
\usepackage{epsfig}
\usepackage{amssymb}
\usepackage{amsmath}
\usepackage{latexsym}
\usepackage{verbatim}

\begin{document}

\def\End{{\sf End}\;}
\let\cd=\cdot

\def\ba{{\bf a}}
\def\bb{{\bf b}}
\def\bc{{\bf c}}
\def\bd{{\bf d}}

\def\CSP{{\rm CSP}}
\def\ECSP{{\rm ECSP}}
\def\SCSP{{\rm SCSP}}
\def\CQE{{\rm CQE}}
\def\ALL{-}

\let\sse=\subseteq
\def\zd{,\ldots,}
\def\rel{R}
\def\tw{\mbox{tw}}
\def\ar{\mbox{arity}}

\def\lb{$\linebreak$}
\let\al=\alpha
\let\gm=\gamma
\let\dl=\delta
\let\ld=\lambda
\let\vf=\varphi
\let\vr=\varrho


\title[Enumerating Homomorphisms]{Enumerating Homomorphisms}
\thanks{The second author is supported by the MCyT through grants
TIN2006-15387-C03-03 and TIN2007-68005-C04-03, and the program Jos\'e
Castillejo. Research of the fourth author is
    supported by the Magyary Zolt\'an Fels\H ooktat\'asi K\"ozalap\'\i
    tv\'any and the Hungarian National Research Fund (Grant Number
    OTKA 67651).}
\author[Andrei]{A.Bulatov}{Andrei A.\ Bulatov}
\address{School of Computing Science,
Simon Fraser University,
Burnaby, Canada}
\email{abulatov@cs.sfu.ca}

\author[Victor]{V.Dalmau}{V\'{\i}ctor\ Dalmau}
\address{Department of Information and Communication Technologies,
Universitat Pompeu Fabra,
Barcelona, Spain}
\email{victor.dalmau@tecn.upf.es}

\author[Martin]{M.Grohe}{Martin Grohe}
\address{Institut f\"ur Informatik,
Humboldt-Universit\"at,
Berlin, Germany}
\email{grohe@informatik.hu-berlin.de}

\author[Daniel]{D.Marx}{D\'aniel Marx}
\address{Department of Computer Science
and Information Theory,
Budapest University of Technology and Economics,
Budapest, Hungary}
\email{dmarx@cs.bme.hu}

\begin{abstract}
  The homomorphism problem for relational structures is an abstract
  way of formulating constraint satisfaction problems (CSP) and
  various problems in database theory. The decision version of the
  homomorphism problem received a lot of attention in literature; in
  particular, the way the graph-theoretical structure of the variables
  and constraints influences the complexity of the problem is
  intensively studied.  Here we
study the problem of
  enumerating all the solutions with polynomial delay from a similar
  point of view. It turns out that the enumeration problem behaves
  very differently from the decision version. We give evidence that it
  is unlikely that a characterization result similar to the decision
  version can be obtained. Nevertheless, we show nontrivial cases
  where enumeration can be done with polynomial delay.
\end{abstract}

\maketitle

\def\zA{\mathbb{A}}
\def\zB{\mathbb{B}}
\def\zC{\mathbb{C}}
\def\zY{\mathbb{Y}}
\def\nat{\mathbb{N}}
\def\cC{\mathcal{C}}
\def\cA{\mathcal{A}}
\def\cB{\mathcal{B}}
\def\cD{\mathcal{D}}
\def\cS{\mathcal{S}}

\section{Introduction}
Constraint satisfaction problems (CSP) form a rich class of
algorithmic problems with applications in many areas of computer
science. We only mention database systems, where CSPs appear in the
guise of the conjunctive query containment problem and the closely
related problem of evaluating conjunctive queries. It has been
observed by Feder and Vardi~\cite{Feder98:monotone} that as abstract problems, CSPs
are homomorphism problems for relational
structures. Algorithms for and the complexity of constraint
satisfaction problems have been intensely studied
(e.g.~\cite{Hell90:h-coloring,Bulatov06:3-element,Barto08:graphs,bodkar08}),
not only for the standard decision problems but also
optimization versions (e.g.~\cite{aus07,jonklakro06,jonkro07,rag08})
and counting versions (e.g. \cite{bul08,buldal03,bulgro05,dyegolpat07}) of CSPs.
\nocite{Schnoor07:enumerating}

In this paper we study the \emph{CSP enumeration problem}, that is,
problem of computing all solutions for a given CSP instance. More
specifically, we are interested in the question which structural
restrictions on CSP instances guarantee tractable enumeration
problems. ``Structural restrictions'' are restrictions on the
structure induced by the constraints on the variables. Example of
structural restrictions is ``every variable occurs in at
most 5 constraints'' or ``the constraints form an acyclic
hypergraph.\footnote{The other type of restrictions studied
  in the literature on CSP are ``constraint language restrictions'',
  that is, restrictions on the structure imposed by the constraint
  relations on the values. An example of a constraint language
  restriction is ``all clauses of a SAT instance, viewed as a Boolean
  CSP, are Horn clauses''.}''
This can most easily be made precise if we
view CSPs as homomorphism problems: Given two relational structures
$\zA,\zB$, decide if there is a homomorphism from $\zA$ to $\zB$. Here the
elements of the structure $\zA$ correspond to the variables of the CSP
and the elements of the structure $\zB$ correspond to the possible
values. Structural restrictions are restrictions on the structure
$\zA$. If $\mathcal A$ is a class of structures, then $\CSP(\mathcal
A,\ALL)$ denotes the restriction of the
general CSP (or homomorphism problem) where the ``left hand side''
input structure $\zA$ is taken from the class $\mathcal
A$. $\ECSP(\mathcal A,\ALL)$ denotes the corresponding enumeration
problem: Given two relational structures $\zA\in\mathcal A$ and $\zB$,
compute the set of all homomorphisms from $\zA$ to $\zB$.  The enumeration
problem is of particular interest in the database context, where we
are usually not only interested in the question of whether the answer
to a query is nonempty, but want to compute all tuples in the
answer. We will also briefly discuss the corresponding \emph{search}
problem: Find a solution if one exists, denoted $\SCSP(\mathcal
A,\ALL)$.

It has been shown in \cite{atsgromar08} that $\ECSP(\mathcal A,\ALL)$ can
be solved in polynomial time if and only if the number of solutions
(that is, homomorphisms) for all instances is polynomially bounded in
terms of the input size and that this is the case if and only if the
structures in the class $\mathcal A$ have bounded fractional edge
cover number. However, usually we cannot expect the number of solutions
to be polynomial. In this case, we may ask which conditions on
$\mathcal A$ guarantee that $\ECSP(\mathcal A,\ALL)$ has a polynomial
delay algorithm. A \emph{polynomial delay algorithm} for an
enumeration problem is required to produce the first solution in
polynomial time and then iteratively compute all solutions (each
solution only once),
leaving
only polynomial time between two successive solutions. In particular,
this guarantees that the algorithms computes all solutions in
\emph{polynomial total time}, that is, in time polynomial in the input
size plus output size.

It is easy to see that $\ECSP(\mathcal A,\ALL)$ has a polynomial
delay algorithm if the class $\mathcal A$ has bounded tree width.
It is also easy to see that there are classes $\mathcal A$ of
unbounded tree width such that $\ECSP(\mathcal A,\ALL)$ has a
polynomial delay algorithm. It follows from our results that
examples of such classes are the class of all grids  or the class
of all complete graphs with a loop on every vertex. 
It is known that the decision problem $\CSP(\mathcal A,\ALL)$
is in polynomial time if and only if the cores of the structures in
$\mathcal A$ have bounded tree width \cite{Grohe03:other-side}
(provided the arity of
the constraints is bounded, and under some reasonable complexity
theoretic assumptions). A \emph{core} of a relational structure
$\mathcal A$ is a minimal substructure $\mathcal A'\subseteq\mathcal
A$ such that there is a homomorphism from $\mathcal A$ to $\mathcal
A'$; minimality is with respect to inclusion. It is easy to see that
all cores of a structure are isomorphic. Hence we usually speak of
``the'' core of a structure. Note that the core of a grid (and of any
other bipartite graph with at least one edge) is a single edge, and
the core of a complete graph with all loops present (and of any other
graph with a loop) is a single vertex with a loop on it. The core of a
complete graph with no loops is the graph itself.  As a polynomial
delay algorithm for an enumeration algorithms yields a polynomial time
algorithm for the corresponding decision problem, it follows that
$\ECSP(\mathcal A,\ALL)$ can only have a polynomial delay algorithm if
the cores of the structures in $\mathcal A$ have bounded tree width.
Unfortunately, there are examples of classes $\mathcal A$ that have
cores of bounded tree width, but for which $\ECSP(\mathcal A,\ALL)$
has no polynomial delay algorithm unless $\textup{P}=\textup{NP}$ (see
Example~\ref{exa:second-solution}).

Our main algorithmic results show that $\ECSP(\mathcal A,\ALL)$ has a
polynomial delay algorithm if the cores of the structures in $\mathcal
A$ have bounded tree width and if, in addition, they can be reached in
a sequence of ``small steps.'' An \emph{endomorphism} of a structure
is a homomorphism of a structure to itself. A \emph{retraction} is an
endomorphism that is the identity mapping on its image. Every
structure has a retraction to its core. However, in
general,
the only way to map a
structure to its core may be by collapsing the whole structure at
once. As an example, consider a path with a loop on both
endpoints. The core consists of a single vertex with a loop. (More
precisely, the two cores are the two endpoints with their loops.) The
only endomorphism of this structure to a proper substructure maps the
whole structure to its core. Compare this with a path that only has a
loop on one endpoint. Again, the core is a single vertex with a loop,
but now we can reach the core by a sequence of retractions, mapping a
path of length $n$ to a subpath of length $n-1$ and then to a subpath
of length $n-2$ et cetera. We prove that if $\mathcal A$ is a class of
structures whose cores have bounded tree width and can be reached
by a sequence of retractions each of which only moves a bounded number of
vertices, then $\ECSP(\mathcal A,\ALL)$ has a polynomial delay algorithm.

We also consider more general sequences of retractions or endomorphism
from a structure to its core.  We say that a sequence of endomorphisms
from a structure $\zA_0$ to a substructure $\zA_1\subset \zA_0$, from $\zA_1$
to a substructure $\zA_2$, \ldots, to a structure $\zA_n$ has
\emph{bounded width} if $\zA_n$ and, for each $i\le n$, the ``difference
between $\zA_i$ and $\zA_{i-1}$'' has bounded tree width. We prove that
if we are given a sequence of endomorphisms of bounded width together
with the input structure $\zA$, then we can compute all solutions by a
polynomial delay algorithm. Unfortunately, in general we cannot
compute such a sequence of endomorphisms efficiently. We prove that
even for width $1$ it is NP-complete to decide whether such a sequence
exists.

Finally, we remark that our results are far from giving a complete
classification of the classes $\mathcal A$ for which $\ECSP(\mathcal
A,\ALL)$ has a polynomial delay algorithm and those classes for which it
does not. Indeed, we show that it will be difficult to obtain such a
classification, because such a classification would imply a solution
to the notoriously open \emph{CSP dichotomy conjecture} of Feder and
Vardi~\cite{Feder98:monotone} (see Section~\ref{sec:tract} for details).

Due to space restrictions several proofs are omitted.

\section{Preliminaries}
\label{sec:preliminaries}

\paragraph{\bf Relational structures.}

A \emph{vocabulary} $\tau$ is a finite set of \emph{relation symbols} of specified
arities. A \emph{relational structure} $\zA$ over $\tau$ consists of a finite
set $A$ called the \emph{universe} of $\zA$ and for each relation symbol
$\rel\in\tau$, say, of arity $r$, an $r$-ary relation $\rel^\zA\sse
A^r$. Note that we require vocabularies and structures to be finite. A
structure $\zA$ is a \emph{substructure} of a structure $\zB$ if
$A\sse B$ and $\rel^\zA\sse \rel^\zB$ for all $R\in\tau$. We write
$\zA\sse\zB$ to denote that $\zA$ is a substructure of $\zB$ and
$\zA\subset\zB$ to denote that $\zA$ is a \emph{proper} substructure of
$\zB$, that is, $\zA\sse \zB$ and $\zA\ne\zB$. A substructure $\zA\sse
\zB$ is \emph{induced} if for all $\rel\in\tau$, say, of arity $r$, we have
$\rel^\zA = \rel^\zB\cap A^r$. For a subset $A\sse B$, we write
$\zB[A]$ to denote the induced substructure of $\zB$ with universe
$A$.

\paragraph{\bf Homomorphisms.}
We often abbreviate tuples $(a_1\zd a_k)$ by $\ba$. If $f$ is a
mapping whose domain contains $a_1\zd a_k$ we write $f(\ba)$ to
abbreviate $(f(a_1)\zd f(a_k))$. A \emph{homomorphism} from a relational
structure $\zA$ to a relational structure $\zB$ is a mapping $\vf :
A\to B$ such that for all $\rel\in\tau$ and all tuples
$\ba\in\rel^\zA$ we have $\vf(\ba)\in\rel^\zB$. A \emph{partial
  homomorphism} on $C\sse A$ to $\zB$ is a homomorphism of $\zA[C]$ to
$\zB$. It is sometimes useful when designing examples to exclude
certain homomorphisms or endomorphisms. The simplest way to do that is
to use unary relations. For example, if $\rel$ is a unary relation and
$(a)\in\rel^\zA$ we say that $a$ \emph{has color} $\rel$. Now if $b\in
B$ does not have color $\rel$ then no homomorphism from $\zA$ to $\zB$
maps $a$ to $b$.

Two structures $\zA$ and $\zB$ are \emph{homomorphically equivalent}
if there is a homomorphism from $\zA$ to $\zB$ and
also a homomorphism from $\zB$ to $\zA$. Note that if structures $\zA$
and $\zA'$ are homomorphically equivalent, then
for every structure $\zB$ there is a homomorphism from $\zA$ to $\zB$
if and only if there is a homomorphism from
$\zA'$ to $\zB$; in other words: the instances $(\zA, \zB)$ and
$(\zA', \zB)$ of the decision CSP are equivalent.
However, the two instances may have vastly different sizes, and the
complexity of solving the search and enumeration problems for them can
also be quite different. Homomorphic equivalence is closely related to
the concept of the core of a structure: A structure $\zA$ is
a \emph{core} if there is no homomorphism from $\zA$ to a proper
substructure of $\zA$. A core of a structure $\zA$ is a
substructure $\zA'\sse \zA$ such that there is a homomorphism from
$\zA$ to $\zA'$ and $\zA'$ is a core. Obviously, every
core of a structure is homomorphically equivalent to the structure. We
observe another basic fact about cores:

\begin{obs}\label{obs:cores}
Let $\zA$ and $\zB$ be homomorphically equivalent structures, and let
$\zA'$ and $\zB'$ be cores of $\zA$ and $\zB$, respectively. Then
$\zA'$ and $\zB'$ are isomorphic. In particular, all cores of a
structure $\zA$ are isomorphic. Therefore, we often speak of
\emph{the} core of $\zA$.
\end{obs}

\begin{obs}
It is easy to see that it is NP-hard to decide, given structures
$\zA\sse \zB$, whether $\zA$ is isomorphic
to the core of $\zB$. (For an arbitrary graph $G$, let $\zA$ be a
triangle and $\zB$ the disjoint union of $G$ with $\zA$. Then
$\zA$ is a core of $\zB$ if and only if $G$ is 3-colorable.) Hell and
Ne\v set\v ril \cite{Hell92:core} proved that it is co-NP-complete to
decide whether a graph is a core.
\end{obs}

\paragraph{\bf Tree decompositions.}
A \emph{tree decomposition} of a graph $G$ is a pair $(T,B)$, where
$T$ is a tree and $B$ is a mapping that associates with
every node $t\in V(T)$ a set $B_t\sse V(G)$ such that
(1) for every $v\in V(G)$ the set $\{t\in V(T) | v\in  B_t\}$ is connected
in $T$, and (2)
for every $e\in E(G)$ there is a $t\in V(T)$ such that $e\sse B_t$.
The sets $B_t$, for $t\in V(T)$, are called the
\emph{bags} of the decomposition. It is sometimes convenient to have
the tree $T$ in a tree decomposition rooted;
we always assume it is. The \emph{width} of a tree decomposition
$(T,B)$ is $\max\{|B_t |\mid t\in V(T)\}-1$. The \emph{tree width} of a
graph $G$, denoted by $\tw(G)$, is the minimum of the widths of all
tree decompositions of $G$.

We need to transfer some of the notions of graph theory to arbitrary
relational structures. The
\emph{Gaifman graph} (also known as \emph{primal graph}) of a
relational structure $\zA$ with vocabulary $\tau$ is the graph
$G(\zA)$ with vertex set $\zA$ and an edge between $a$ and $b$ if
$a\ne b$ and there is a relation symbol $\rel\in\tau$, say, of arity
$r$, and a tuple $(a_1\zd a_r)\in\rel^\zA$ such that $a, b\in \{a_1\zd
a_r\}$. We can now transfer graph-theoretic notions to relational
structures. In particular, a subset $B\sse A$ is \emph{connected} in a
structure $\zA$ if it is connected in $G(\zA)$. A \emph{tree
  decomposition} of a structure $\zA$ can simply be defined to be a
tree-decomposition of $G(\zA)$. Equivalently, a tree decomposition of
$\zA$ can be defined directly by replacing the second condition in the
definition of tree decompositions of graphs by (2')
for every $\rel\in\tau$ and $(a_1\zd a_r)\in\rel^\zA$ there is a $t\in
V(T)$ such that $\{a_1\zd a_r\}\sse B_t$.
A class $\cC$ of structures has \emph{bounded tree width} if there is a
$w\in\nat$ such that $\tw(\zA)\le w$ for all $\zA\in \cC$. A
class $\cC$ of structures has \emph{bounded tree width modulo
  homomorphic equivalence} if there is a $w\in \nat$ such that
every $\zA\in \cC$ is homomorphically equivalent to a structure of
tree width at most w.

\begin{obs}\label{obs:width-cores}
A structure $\zA$ is homomorphically equivalent to a structure of
tree width at most $w$ if and only if the core of $\zA$ has tree width
at most $w$.
\end{obs}

\paragraph{\bf The Constraint Satisfaction Problem.}
For two classes $\cA$ and $\cB$ of structures, the \emph{Constraint
  Satisfaction Problem}, $\CSP(\cA,\cB)$, is the following problem:

\vskip1mm

\noindent
\begin{center}
\fbox{\parbox{0.6\linewidth}{
$\CSP(\cA,\cB)$

\emph{Instance:} \ $\zA\in\cA$, $\zB\in\cB$

\emph{Problem:} \ Decide if there is a homomorphism from $\zA$ to
$\zB$.}}
\end{center}

\vskip2mm

The CSP is a decision problem. The variation of it we study in this
paper is the following enumeration problem:

\vskip2mm

\noindent
\begin{center}
\fbox{\parbox{0.6\linewidth}{
$\ECSP(\cA,\cB)$

\emph{Instance:} \ $\zA\in\cA$, $\zB\in\cB$

\emph{Problem:} \ Output all the homomorphisms from $\zA$ to $\zB$.}}
\end{center}

\vskip2mm

We shall also refer to the search problem, $\SCSP(\cA,\cB)$, in which
the goal is to find one solution to a CSP-instance or output `no' if a
solution does not exists.

If one of the classes $\cA$, $\cB$ is the class of all finite
structures, then we denote the corresponding CSPs by $\CSP(\cA,\ALL)$,
$\CSP(\ALL,\cB)$ (respectively, $\ECSP(\cA,\ALL)$, $\ECSP(\ALL,\cB)$,
$\SCSP(\cA,\ALL)$, $\SCSP(\ALL,\cB)$).

The decision CSP has been intensely studied. If a class $\cC$ of
structures has bounded arity then $\CSP(\cC,\ALL)$ is solvable in
polynomial time if and only if $\cC$ has bounded tree width modulo
homomorphic equivalence \cite{Grohe03:other-side}. If the arity of
$\cC$ is not bounded, several quite general conditions on a class of
structures have been identified that guarantee polynomial time
solvability of $\CSP(\cC,\ALL)$, see,
e.g.\cite{Gottlob02:hypertree,chen/dalmau:05,Grohe06:fractional}.
Problems of the form $\CSP(\ALL,\cC)$ have been studied mostly in
the case when $\cC$ is 1-element. Problems of this type are
sometimes referred to as \emph{non-uniform}. It is conjectured that
every non-uniform problem is either solvable in polynomial time or
NP-complete (the so-called \emph{Dichotomy Conjecture})
\cite{Feder98:monotone}. Although this conjecture is proved in
several particular cases
\cite{Hell90:h-coloring,Bulatov03:conservative,Bulatov06:3-element,Barto08:graphs},
in its general form it is believed to be very difficult.

A search CSP is clearly no easier than the corresponding decision
problem. While any non-uniform search problem $\SCSP(\ALL,\cC)$ is
polynomial time reducible to its decision version $\CSP(\ALL,\cC)$
\cite{Bulatov05:classifying}, nothing is known about the complexity
of search problems $\SCSP(\cC,\ALL)$ except the result we state in
Section~\ref{sec:tract}. Paper \cite{Schnoor07:enumerating} provides
some initial results on the complexity of non-uniform enumerating
problems.

\section{Tractable structures for enumeration}\label{sec:tract}

Since even an easy CSP may have exponentially many solutions, the
model of choice for `easy' enumeration problems is algorithms with
polynomial delay \cite{Johnson88:generating}. An algorithm Alg is said
to solve a CSP \emph{with polynomial delay} (\emph{WPD} for short) if
there is a polynomial $p(n)$ such that, for every instance of size
$n$, Alg outputs `no' in a time bounded by $p(n)$ if there is no solution,
otherwise it generates
all solutions to the instance such that no solution is output twice,
the first solution is output after at most $p(n)$ steps after the
computation starts, and time between outputting two consequent
solutions does not exceed $p(n)$.

If a class of relational structures $\cC$ has bounded arity, the
aforementioned result of Grohe \cite{Grohe03:other-side} imposes
strong restrictions on enumeration problems solvable WPD.

\begin{obs}\label{obs:bw-core}
If a class of relational structures $\cC$ with bounded arity does not have bounded
tree width modulo homomorphic equivalence, then $\ECSP(\cC,\ALL)$ is not WPD, unless P$=$NP.
\end{obs}

Unlike for the decision version, the converse is not true: bounded
tree width modulo homomorphic equivalence does not imply enumerability WPD.

\begin{example}\label{exa:second-solution}\rm
  Let $\zA_k$ be the disjoint union of a $k$-clique and a loop and let
  $\cA=\{\zA_k \;|\; k\ge 1\}$. Clearly, the core of each graph in
  $\cA$ has bounded tree width (in fact, it is a single element),
  hence $\CSP(\cA,\ALL)$ is
  polynomial-time solvable. For an arbitrary graph $\zB$ without loops, let $\zB'$ be
  the disjoint union of $\zB$ and a loop. It is clear that there is
  always a trivial homomorphism from $\zA_k$ (for any $k\ge 1$) to
  $\zB'$ that maps everything into the loop.  There exist
  homomorphisms different from the trivial one if and only if $\zB$
  contains a $k$-clique. Thus if we are able to check in polynomial
  time whether there is a second homomorphism, then we are able to
  test if $\zB$ has a $k$-clique. Therefore, although
  $\CSP(\cA,\ALL)$ and $\SCSP(\cA,\ALL)$ are polynomial-time solvable,
  a WPD enumeration algorithm
  for $\ECSP(\cA,\ALL)$ would imply $\text{P}=\text{NP}$.
\end{example}

It is not difficult to show that $\ECSP(\cC,\ALL)$ is enumerable WPD
if $\cC$ has bounded tree width. For space restrictions we do not
include a direct proof and instead we derive it from a more general
result in Section~\ref{sec:sequence}. Thus enumerability WPD has a
different tractability criterion than the decision version, and this
criterion lies somewhere between bounded tree width and bounded tree
width modulo homomorphic equivalence. Thus in order to ensure that
the solutions can be enumerated WPD, we have to make further
restrictions on the way the structure can be mapped to its bounded
tree width core. The main new definition of the paper requires that
the core is reached by ``small steps'':

Let $\zA$ be a relational structure with universe $A$. We
say that $\zA$ \emph{has a sequence of endomorphisms of width $k$} if there are
subsets $A = A_0\supset A_1\supset\ldots\supset
A_n\ne\emptyset$
and homomorphisms $\vf_1,\ldots,\vf_n$ such that
\begin{enumerate}
\item
$\vf_i$ is a homomorphism from $\zA[A_{i-1}]$ to $\zA[A_i]$,
\item
$\vf_i(A_{i-1}) = A_i$ for $1\le i\le n$;
\item
if $G$ is the primal graph of $\zA$, then the tree width of $G[A_i \setminus
  A_{i+1}]$ is at most $k$ for every $0\le i< n$;
\item
the structure induced by $A_n$
has
tree width at most $k$.
\end{enumerate}

In Section~\ref{sec:sequence}, we show that enumeration for $(\zA,\zB)$ can be done
WPD if a sequence of bounded width endomorphisms for $\zA$ is
given in the input. Unfortunately, we cannot claim that
$\ECSP(\cA,\ALL)$ can be done WPD if every structure in $\cA$ has such a
sequence, since we do not know how to find such sequences
efficiently. In fact, as we show in Section~\ref{sec:hardness}, it is
hard to check if a width-1 sequence exists for a given
structure. Furthermore, we show a class $\cA$ where every structure
has a width-2 sequence, but $\ECSP(\cA,\ALL)$ cannot be done WPD,
unless $\text{P}=\text{NP}$. This means that it is not possible to get around the
problem of not being able to find the sequences (for example, by
finding sequences with somewhat larger width or by constructing the
sequence during the enumeration).

Thus having a bounded width sequence of endomorphisms is not the
right tractability criterion. We then investigate a more restrictive
notion, where the bound is not on the tree width of the difference
of the layers but on the number of elements in the differences.
However, in the rest of the section, we give evidence that
enumeration problems solvable WPD cannot be characterized in simple
terms relying on tree width. For instance, a description of search
problems solvable in polynomial time would imply a description of
non-uniform decision problems solvable in polynomial time.
%
%
This is shown via an
analogous result for the search version of the problem, which might
be of independent interest. By $\zA\oplus\zB$ we
denote the disjoint union of relational structures $\zA$ and~$\zB$.

\begin{lemma}\label{lem:enum-vs-non-uniform}
Let $\zB$ be a relational structure, which is a core, and let
$\cC_\zB$ be $\{\zA\oplus\zB\mid
\zA\to\zB\}$. Then $\CSP(\ALL,\zB)$ is
solvable in polynomial time if and only if so is the problem $\SCSP(\cC_\zB,\ALL)$.
\end{lemma}

\begin{proof}
If the decision problem $\CSP(\ALL,\zB)$ is solvable in polynomial
time we can construct an
algorithm that given an instance $(\zA,\zC)$ of $\CSP(\cC_\zB,\ALL)$
computes a solution in polynomial time. Indeed, as $\CSP(\ALL,\zB)$ is
solvable in polynomial time by the aforementioned result of
\cite{Bulatov05:classifying} it is also polynomial time to find a
homomorphism from a given structure to $\zB$ provided one exists. If
$\zA\in\cC_\zB$ such a homomorphism $\vf$ exists by
the definition of $\cC_\zB$. So our algorithms, first, finds some
homomorphism $\vf$. Then it decides
by
brute force whether or not there exists a homomorphism $\vf'$ from
$\zB$ to $\zC$ (note that this can be done in polynomial time for
every fixed $\zB$). If such a homomorphism does not exist then we can
certainly guarantee that there is no homomorphism from $\zA$ to
$\zC$. Otherwise we obtain a required homomorphism $\psi$ as follows:
Let $\psi(a)=\vf'(a)$ for $a\in\zB$, and $\psi(a)=\vf'\circ\vf(a)$ for
$a\in\zA$.

Conversely, assume that we have an algorithm Alg that finds a solution
of any instance of $\CSP(\cC_\zB,\ALL)$ in polynomial time, say,
$p(n)$. We construct from it an algorithm that solves
$\CSP(\ALL,\zB)$. Given an instance $(\zA,\zB)$ of $\CSP(\ALL,\zB)$ we call algorithm
Alg with input $\zA\oplus\zB$ and $\zB$. Additionally we count the
number of steps performed by Alg in such a way that we stop if Alg has
not finished in $p(n)$ steps. If Alg produces a correct answer then we
have to be able to obtain from it a homomorphism from $\zA$ to
$\zB$. If Alg's answer is not correct or the clock reaches $p(n)$
steps we know that Alg failed. The only possible reason for that is
that $\zA\oplus\zB$ does not belong to $\cC_\zB$, which implies that
$\zA$ is not homomorphic to $\zB$.
\end{proof}

In what follows we transfer this result to enumeration problems.
Let
$\cA$ be a class of relational structures. The class $\cA'$
consists of all structures built as follows: Take $\zA\in\cA$ and add
to it $|\zA|$ independent vertices.

\begin{lemma}\label{lem:search-enumeration}
Let $\cA$ be a class of relational structures. Then $\SCSP(\cA,\ALL)$ is
solvable in polynomial time if and only if $\ECSP(\cA',\ALL)$ is
solvable WPD.
\end{lemma}

\begin{proof}
If $\ECSP(\cA,\ALL)$ is enumerable WPD, then for any
structure $\zA'\in\cA'$ it takes time polynomial in $|\zA'|$ to find
the first solution. Since $\zA'$ is only twice of the size of the
corresponding structure $\zA$, it takes only polynomial time to solve
$\SCSP(\cA,\ALL)$.

Conversely, given a structure $\zA'=\zA\cup I\in\cA'$, where
$\zA\in\cA$ and $I$ is the set of independent elements, and any
structure $\zB$. The first homomorphism from $\zA'$ to $\zB$ can be
found in polynomial time, since $\SCSP(\cA,\ALL)$ is polynomial time
solvable and the independent vertices can be mapped arbitrarily. Let
the restriction of this homomorphism onto $\zA$ be $\vf$. Then while
enumerating all possible $|\zB|^{|\zA|}$ extensions of $\vf$ we buy
enough time to enumerate all homomorphisms from $\zA$ to $\zB$ using
brute force.
\end{proof}

\vspace*{-1mm}

\section{Sequence of bounded width endomorphisms}
\label{sec:sequence}

In this section we show that for every fixed $k$, all the homomorphisms
from $\zA$ to $\zB$ can be enumerated with polynomial delay if a sequence of
width $k$ endomorphisms of $\zA$ is given in the input. Given a
sequence $A_0,\dots,A_n$ and $\vf_1$, $\dots$, $\vf_n$ as in
the definition of a sequence of width~$k$ endomorphisms,
we denote $\zA[A_i]$ by $\zA_i$.

We will enumerate the homomorphisms from $\zA$ to $\zB$ by first enumerating
the homomorphisms from $\zA_n$, $\zA_{n-1}$, $\dots$ to $\zB$ and
 then transforming them to homomorphisms
from $\zA$ to $\zB$ using the homomorphisms $\vf_i$. We obtain the
homomorphisms from $\zA_i$ by
extending the homomorphism from $\zA_{i+1}$ to the set $A_i\setminus
A_{i+1}$; Lemma~\ref{lem:chain-extention} below will be useful for
this purpose. In order to avoid producing a homomorphism multiple
times, we need a delicate classification
(see definitions of elementary homomorphisms and of the index of a homomorphism).

\begin{lemma}\label{lem:chain-extention}
Let $\zA,\zB$ be relational structures and $X_1 \subseteq X_2\subseteq A$
subsets, and let $g_0$ be a
homomorphism from $\zA[X_1]$ to $B$. For every fixed $k$, there is a polynomial-time
algorithm {\sc Homomorphism-Ext}$(\zA,\zB,X_1,X_2, g_0)$ that decides
whether $g_0$ can be extended to a homomorphism from $\zA[X_2]$ to
$\zB$, if the tree width of induced subgraph $G[X_2\setminus X_1]$ of the Gaifman graph of $\zA$ is at most $k$.
\end{lemma}

The \emph{index} of a homomorphism $\vf$ from $\zA$ to $\zB$ is the
largest $t$ such that $\vf$
can be written as $\vf = \psi \circ \vf_t \circ\ldots\circ\vf_1$ for
some homomorphism $\psi$ from
$\zA_t$ to $\zB$. In particular, if $\vf$ cannot be written as $\vf =
\psi\circ\vf_1$, then the
index of $\vf$ is 0.
Observe that if the index of $\vf$ is at least $t$, then
there is a unique $\psi$ such that $\vf = \psi \circ \vf_t
\circ\ldots\circ\vf_1$: This follows
from the fact that $\vf_t \circ\ldots\circ\vf_1$ is a surjective mapping from $A$ to
$A_t$, thus if $\psi'$ and $\psi''$ differ on $A_t$, then $\psi' \circ \vf_t
\circ\ldots\circ\vf_1$ and $\psi'' \circ \vf_t
\circ\ldots\circ\vf_1$ differ on $A$.
A homomorphism $\psi$ from $\zA_t$
to $\zB$ is \emph{elementary}, if it cannot be written as $\psi = \psi'\circ
\vf_{t+1}$. A homomorphism is {\em reducible} if it is not elementary.

\begin{lemma}\label{lem:elementary-index}
If a homomorphism $\psi$ from $\zA_t$ to $\zB$ is elementary,
then $\vf = \psi \circ \vf_t \circ\ldots\circ\vf_1$ has index exactly
$t$. Conversely, if homomorphism $\vf$ from $\zA$ to $\zB$ has index
$t$ and can be written as $\vf = \psi \circ \vf_t
\circ\ldots\circ\vf_1$, then the homomorphism $\psi$ from $\zA_t$ to $\zB$ is
elementary.
\end{lemma}

Lemma~\ref{lem:elementary-index} suggests a
way of enumerating all the homomorphisms from $\zA$ to $\zB$: for $t =
0\zd n$, we enumerate all the elementary homomorphisms from $\zA_t$ to $\zB$,
and for each such homomorphism $\psi$, we compute $\vf = \psi \circ
\vf_t \circ\ldots\circ\vf_1$. To this end, we need the following
characterization of elementary homomorphisms:

\begin{lemma}\label{lem:hom-elementary}
A homomorphism $\psi$ from $\zA_t$ to $\zB$ is reducible
if and only if
\begin{enumerate}
\item[(1)] $\psi(x)=\psi(y)$ for every $x, y\in A_t$ with
$\vf_{t+1}(x) = \vf_{t+1}(y)$, i.e., for every $z\in A_{t+1}$,
$\psi(x)$ has the same value $b_z$ for every $x$ with
$\vf_{t+1}(x)=z$, and
\item[(2)] the mapping defined by $\psi'(z):=b_z$ is a homomorphism
  from $\zA_{t+1}$ to $\zB$.
\end{enumerate}
\end{lemma}

Lemma~\ref{lem:hom-elementary} gives a way of testing in
polynomial time whether a given homomorphism $\psi$ is elementary: we
have to test whether one of the two conditions are violated.  We state
this in a more general form: we can test in polynomial time whether a
partial mapping $g_0$ can be extended to an elementary homomorphism
$\psi$, if the structure induced by the elements where $g_0$ is not
defined has bounded tree width. We fix values every possible way in
which the conditions of Lemma~\ref{lem:hom-elementary} can be
violated and use {\sc Homomorphism-Ext} to check whether there is an
extension compatible with this choice.  In order to efficiently
enumerate all the possible violations of the second condition, the
following definition is needed:

Given a relation $\rel^\zB$ of arity $r$, a \emph{bad
prefix} is a tuple $(b_1\zd b_s)\in B^s$ with $s\le r$ such that
\begin{enumerate}
\item
there is no tuple $(b_1\zd b_s, b_{s+1}\zd b_r)\in\rel^\zB$ for any
$b_{s+1}\zd b_r\in B$, and
\item
there is a tuple $(b_1\zd b_{s-1}, c_s, c_{s+1}\zd c_r)\in \rel^\zB$
for some $c_t\zd c_r\in B$.
\end{enumerate}

If $(b_1\zd b_r)\not\in\rel^\zB$, then there is
a unique $1\le s\le r$ such that the tuple $(b_1\zd
b_s)$ is a bad prefix: there has to be an $s$ such that $(b_1\zd b_s)$
cannot be extended to a tuple of $\rel^\zB$, but $(b_1\zd b_{s-1})$
can.

\begin{lemma}\label{lem:few-bad-prefixes}
The relation $\rel^\zB$ has at most $|\rel^\zB|\cdot (|B|-1)\cdot r$
bad prefixes, where $r$ is the arity of the relation.
\end{lemma}

\begin{lemma}\label{lem:elementary-ext}
Let $X$ be a subset of $A_t$ and let $g_0$ be a
mapping from $X$ to $B$. For every fixed $k$, there is a polynomial-time
algorithm {\sc Elementary-Ext}$(t,X, g_0)$ that decides whether $g_0$ can be
extended to an elementary homomorphism from $\zA_t$ to $B$, if the tree width
of the structure induced by $A_t-X$ is at most $k$.
\end{lemma}

We enumerate the elementary homomorphisms in a specific order defined by the
following precedence relation.
Let $\vf$ be an elementary homomorphism from $\zA_i$ to $\zB$ and let
$\psi$ be an elementary homomorphism from $\zA_j$ to $\zB$ for some $j >
i$. Homomorphism $\psi$ is the \emph{parent} of $\vf$ ($\vf$ is a
\emph{child} of $\psi$) if $\vf$ restricted to $A_{i+1}$ can
be written as
$\psi\circ\vf_j\circ\ldots\circ\vf_{i+2}$. \emph{Ancestor} and
\emph{descendant} relations are defined as the reflexive transitive
closure of the parent and child relations, respectively.

Note that an elementary homomorphism from $\zA_i$ to $\zB$ has exactly
one parent for $i < n$ and a homomorphism from
$\zA_n$ to $\zB$ has no parent.
Fix an arbitrary ordering of the elements of
$A$. For $0 \le i\le n$ and $0\le j\le |A_i \setminus A_{i+1}|$, let $A_{i,j}$
be the union of $A_{i+1}$ and the first $j$ elements of $A_i\setminus
A_{i+1}$. Note that $A_{i,0} = A_{i+1}$ and
$A_{i,|A_i\setminus A_{i+1}|} = A_i$.

\begin{lemma}\label{lem:elementary-enum}
Let $\psi$ be a mapping from $A_{i,j}$ to $\zB$
that can be extended to an elementary homomorphism from $\zA_i$ to $\zB$. Assume that
a sequence of width $k$ endomorphisms is given for $\zA$. For every fixed
$k$, there is a polynomial-delay, polynomial-space algorithm
{\sc Elementary-Enum}$(i, j, \psi)$ that enumerates all the elementary
homomorphisms of $\zA_i$ that extends $\psi$ and all the descendants of these
homomorphisms.
\end{lemma}

By calling {\sc Elementary-Enum}$(n, 0, g_0)$ (where $g_0$ is a
trivial mapping from $\emptyset$ to $\zB$),
we can enumerate all the elementary homomorphisms. By the observation
in Lemma~\ref{lem:elementary-index}, this means that we can enumerate
all the homomorphisms from $\zA$ to $\zB$.

\begin{theorem}\label{the:width-k-chain}
For every fixed $k$, there is a
polynomial-delay, polynomial-space algorithm that, given structures $\zA$,
$\zB$, and a sequence of width $k$ endomorphisms of $\zA$, enumerates all the
homomorphisms from $\zA$ to~$\zB$.
\end{theorem}

Theorem~\ref{the:width-k-chain} does not provide a complete
description of classes of structures solvable WPD.

\begin{corollary}\label{cor:unbounded-width}
There is a class $\cA$ of relational structures such that not all structures from $\cA$
have a sequence of width $k$ endomorphisms and $\ECSP(\cA,\ALL)$ is solvable
WPD.
\end{corollary}

\begin{proof}
Let $\cA$ be the class of structures that are the disjoint union of a
loop and a core. Obviously, $\SCSP(\cA,\ALL)$ is polynomial time
solvable. Therefore, by Lemma~\ref{lem:search-enumeration},
$\ECSP(\cA',\ALL)$ is solvable with polynomial delay. However, it is not
hard to see that $\cA'$ does not have a sequence of endomorphisms of bounded tree width.
\end{proof}

Furthermore, as we will see in the next section it is hard, in general, to find
a sequence of bounded width endomorphims. Still, we can find a sequence of endomorphisms
for a structure $\zA$ if we impose two more restrictions on such a sequence.

A retraction $\vf$ of a structure $\zA$ is called a
\emph{$k$-retraction} if at most $k$ nodes change their value
according to $\vf$. A structure is a \emph{$k$-core} if the only
$k$-retraction is the identity. A $k$-core of a structure is any
$k$-core obtained by a sequence of $k$-retractions.

\begin{lemma}\label{lem:k-cores-isomorphic}
All $k$-cores of a structure $\zA$ are isomorphic.
\end{lemma}

Lemma~\ref{lem:k-cores-isomorphic} amounts to say that when searching
for a sequence of $k$-retractions converging to a $k$-core we can use
the greedy approach and include, as the next member of such a
sequence, any $k$-retraction with required properties. With this in
hands we now can apply Theorem~\ref{the:width-k-chain}.

\begin{theorem}\label{the:finite-retractions}
Let $k>0$ be a positive integer and let $\cC$ be a class of structures
such that the $k$-core of every structure in $\cC$ has tree width at
most $k$. Then, the enumeration problem $\ECSP(\cC,\ALL)$ is solvable
WPD.
\end{theorem}

\begin{corollary}
If $\cC$ is a class of structures of bounded tree width then $\ECSP(\cC,\ALL)$ is solvable
WPD.
\end{corollary}

\section{Hardness results}\label{sec:hardness}

The first result of this section shows that finding a sequence of
endomorphisms of bounded width can be difficult even in simplest
cases.

\begin{theorem}\label{the:hardness}
It is NP-complete to decide if a structure has a sequence of
1-width retractions to the core.
\end{theorem}

The second result shows that $\ECSP(\cA,\ALL)$ can be hard even if
every structure in $\cA$ has a sequence of width-2 endomorphisms. Note
that this result is incomparable with Theorem~\ref{the:hardness},
since an enumeration algorithm (in theory) does not necessarily have
to compute an sequence of endomorphisms. We need the following lemma:

\begin{lemma}\label{lem:outerplanar}
If $G$ is a planar graph, then it is possible to find a partition
$(V_1,V_2)$ of its vertices in polynomial time such that $G[V_1]$ and
$G[V_2]$ have tree width at most $2$.
\end{lemma}

\begin{prop}\label{pro:fpt-not-poly}
  There is a class $\cA$ of relational structures
  such that every structure from $\cA$ has a sequence of width 2
  endomorphisms to the core, and such that the problem $\ECSP(\cA,\ALL)$ is
  not solvable WPD, unless $P=NP$.
\end{prop}

\begin{proof}
  Let $\cA$ be a class of graphs built in the following way. Take a
  3-colorable planar graph $G$ and its partition $(V_1,V_2)$ according
  to Lemma~\ref{lem:outerplanar}.  Using colorings we can ensure that
  $G$ is a core. Then we take a disjoint union of this graph with a
  triangle $T$ having all the colors and a copy $G_1$ of $G[V_1]$. Let
  $\zA$ denote the resulting structure.

{\sc Claim 1.}
$\zA$ has a sequence of width-2 endomorphisms.\\[1mm]
\indent
Let $\psi$ be a 3-coloring of $G$ that is a homomorphism into the
triangle, and $\psi'$ the bijective mapping from $G_1$ to $G[V_1]$.
Then $\vf_1$ is defined to act as $\psi$ on $G$, as $\psi'$ on $G'_1$
and identically on $T$. Endomorphism $\vf_2$ is just the 3-coloring of
$G\cup G_1$ induced by $\psi$. The images of $\vf_1$ and $\vf_2$ are
$T\cup G[V_1]$ and $T$, respectively, so all the conditions on a
sequence of width-2 homomorphisms  are easily checkable.

\smallskip

{\sc Claim 2.}
 The {\sc Planar graph 3-coloring problem} is Turing reducible to
 $\ECSP(\cA,\ALL)$.\\[1mm]
\indent
Given a planar graph $G$ we find its partition $(V_1,V_2)$ and create
a structure $\zA$, as described above. Then we apply an algorithm that
enumerates solutions to $\ECSP(\cA,\ALL)$
We may assume that
such an algorithm stops with some time bound regardless whether $G$ is
3-colorable or not. If the algorithm succeeds we can now produce a
3-coloring of $G$.
\end{proof}

\section{Conjunctive queries}\label{sec:conjunctive-queries}

When making a query to a database one usually needs to obtain values
of only those variables (attributes) (s)he is interested in. In terms
of homomorphisms this can be translated as follows: For relational
structures $\zA$, $\zB$, and a subset $Y\sse A$, we aim to list those
mappings from $Y$ to $B$ which can be extended to a full homomorphism
from $\zA$ to $\zB$. In other words, we would like to enumerate all
the mappings from $Y$ to $B$ that arise as the restriction of some
homomorphism from $\zA$ to $\zB$. Clearly, this problem significantly differs from
the regular enumeration problem. A mapping from $Y$ to $B$ can be
extendible to a homomorphism in many ways, possibly superpolynomially
many, and an enumeration algorithm would list all of them. In the
worst case scenario it would list them before turning to the next
partial mapping. If this happens it may destroy polynomiality of the
delay between outputting consecutive solutions.

In this section we treat the {\sc Conjunctive Query Evaluation problem} as follows.

\vskip4mm \noindent
\begin{center}
\fbox{\parbox{0.6\linewidth}{
$\CQE(\cA,\cB)$

\emph{Instance:} \ $\zA\in\cA$, $\zB\in\cB$, $Y\sse A$

\emph{Problem:} \ Output all partial mappings from $Y$ to $B$
extendible to a homomorphism from $\zA$ to $\zB$.}}
\end{center}

\vskip2mm

We present two results, first one of them shows that the problem
$\CQE(\cA,\ALL)$ is WPD when $\cA$ is a class of structures of bounded
tree width, the second one claims that, modulo some complexity assumptions,  in contrast to enumeration
problems this cannot be generalized to structures with $k$-cores of
bounded tree width for $k\ge 2$.

\begin{theorem}\label{the:cqe-btw}
If $\cA$ is a class of structures of bounded width then $\CQE(\cA,\ALL)$ is solvable WPD.
\end{theorem}

\begin{proof}
We use Lemma~\ref{lem:chain-extention} to show that algorithm {\sc
  CQE-Bounded-Width} of Figure~\ref{fig:cqe-btw} does the job. Indeed,
  this algorithms backtracks only if outputs a solution.
\end{proof}

Theorem~\ref{the:cqe-btw} does not generalize to classes of
structures whose $k$-cores have bounded width.

\begin{figure}[b]
\caption{Algorithm {\sc CQE-Bounded-Width}}\label{fig:cqe-btw}
\flushleft
{\it Input:} Relational structures $\zA$, $\zB$, and $Y=\{Y_1\zd Y_{\ell}\}\sse A$\\[1mm]
{\it Output:} A list of mappings $\vf\colon Y\to B$ extendible to a
homomorphism from $\zA$ to $\zB$ \small
\begin{tabbing}
{\it Step 1} \ \ \ \ \ \= {\bf set} $m=0$, $\vf=\emptyset$, $S_i=B$, $i\in[m]$, complete$:=${\bf false}\\
{\it Step 2} \> {\bf while} not complete {\bf do}\\
{\em Step 2.1} \> \ \ \ \ \= {\bf if} $m<\ell$ {\bf then do}\\
{\em Step 2.1.1} \> \> \ \ \ \ \= {\bf search} $S_{m+1}$ until
a $b\in S_{m+1}$ is found such that there exists a homomorphism extending \\
\> \> \> $\vf\cup\{y_{m+1}\rightarrow b \}$ and {\bf remove} all members of $S_{m+1}$ preceding $b$ inclusive\\
{\em Step 2.1.2} \> \> \> {\bf if} such a $b$ exists {\bf then
  set} $\vf:=\vf\cup\{y_{m+1}\rightarrow b\}$, $m:=m+1$\\
{\em Step 2.1.3} \> \> \> {\bf else} \\
{\em Step 2.1.3.1} \> \> \> \ \ \= {\bf if} $m\ne0$ {\bf then set}
$\vf=\vf_{|\{y_1,\dots,y_{m-1}\}}$ and
$S_{m+1}:=B$, $m:=m-1$ \\
{\em Step 2.1.3.2} \> \> \> \> {\bf else set} complete:={\bf true} \\
{\em Step 2.2} \> \> {\bf else then do}\\
{\em Step 2.2.1} \> \> \> {\bf output} $\vf$\\
{\em Step 2.2.2} \> \> \> {\bf set} $\vf:=\vf_{|\{y_1,\dots,y_{m-1}\}\}}$, $m: =\ell-1$\\
\> {\bf endwhile}
\end{tabbing}
\vspace*{-4mm}
\end{figure}

\begin{example}\label{exa:multi-col-clique}\rm
  Recall that the {\sc Multicolored Clique} problem (cf.~\cite{multiple-interval}) is
  formulated as follows: Given a number $k$ and a vertex $k$-colored
  graph, decide if the graph contains a $k$-clique all vertices of
  which are colored different colors. This problem is
  $W[1]$-complete,
 i.e., has no time $f(k)n^c$ algorithm for
  any function $f$ and constant $c$, unless FPT$=W[1]$. We
  reduce this problem to  $\CQE(\cA,\ALL)$ where $\cA$
  is the class of structures whose 2-cores are 2-element described
  below.

  Let us consider relational structures with two binary and two unary
  relations. This structure can be thought of as a graph whose
  vertices and edges have one of the two colors, say, red and blue,
  accordingly to which of the two binary/unary relations they belong
  to. Let $\zA_k$ be the relational structure with universe $\{a_1\zd
  a_k,y_1\zd y_k\}$, where $a_1\zd a_k$ are red while $y_1\zd y_k$ are
  blue. Then $\{a_1\zd a_k\}$ induces a red clique, that is every
  $a_i,a_j$ ($i,j$ are not necessarily different) are connected with a
  red edge, and each $y_i$ is connected to $a_i$ with a
  blue edge. It is not hard to see that every pair of a red and blue
  vertices induces a 2-core of this structure. Set $\cA=\{\zA_k\mid
  k\in\nat\}$.

The reduction of the {\sc Multicolored Clique} problem to
$\CQE(\cA,\ALL)$ goes as follows. Given a $k$-colored graph $G=(V,E)$
whose coloring induces a partition of $V$ into classes $B_1\zd
B_k$. Then we define structures $\zA,\zB$ and a set $Y\sse A$. We set
$\zA=\zA_k$, $Y=\{y_1\zd y_k\}$. Then let $B=V\cup\{b_1\zd b_k\}$, the
elements of $V$ are colored red and the induced substructure $\zB[V]$
is the graph $G$ (without coloring) whose edges are colored also
red. Finally, $b_1\zd b_k$ are made blue and each $b_i$ is connected
with a blue edge with every vertex from $B_i$.

It is not hard to see that any homomorphism maps $\{a_1\zd a_k\}$ to
$V$ and $Y$ to $\{b_1\zd b_k\}$, and that the number of homomorphisms
that do not agree on $Y$ does not exceed $k^k$. Moreover, $G$ contains
a $k$-colored clique if and only if there is a homomorphism from $\zA$
to $\zB$ that maps $Y$ onto $\{b_1\zd b_k\}$. If there existed an
algorithm solving $\CQE(\cA,\ALL)$ WPD, say, time needed to compute
the first and every consequent solution is bounded by a polynomial
$p(n)$, then time needed to list all solutions is at most
$k^kp(n)$. This means that {\sc Multicolored Clique} is FPT, a
contradiction.
\end{example}

\vspace*{-4mm}
\bibliographystyle{plain}

\end{document}